\documentclass[aps,prd,twocolumn,groupedaddress,showpacs,nofootinbib]{revtex4}

\usepackage{color}
\usepackage{graphicx}
\usepackage{dcolumn}
\usepackage{bm}
\usepackage{amssymb,amsmath,latexsym,amsfonts}

\begin{document}

\title{Bose--Einstein Condensation and Symmetry Breaking of a Complex Charged Scalar Field}

\author{Tonatiuh Matos\footnote{Part of the Instituto Avanzado de Cosmolog\'ia (IAC) collaboration
http://www.iac.edu.mx/}} \email{tmatos@fis.cinvestav.mx}
\affiliation{Departamento de F\'isica,\\ Centro de Investigaci\'on y
de Estudios Avanzados del IPN, A.P. 14-740, 07000 M\'exico D.F.,
M\'exico.}

\author{El\'ias Castellanos} \email{ecastellanos@mctp.mx, ecastellanos@fis.cinvestav.mx}
\affiliation{Mesoamerican Centre for Theoretical Physics\\Universidad Aut\'onoma de Chiapas,
 \\Ciudad Universitaria,
Carretera Zapata Km. 4, Real del Bosque (Ter\'an), 29040, Tuxtla Guti\'errez, Chiapas, M\'exico.}
\affiliation{Departamento de F\'isica,\\ Centro de Investigaci\'on y de Estudios Avanzados
del IPN, A.P. 14-740, 07000 M\'exico D.F., M\'exico.}

\author{Abril  Su\'arez\footnote{Part of the Instituto Avanzado de Cosmolog\'ia (IAC) collaboration
http://www.iac.edu.mx/}} \email{asuarez@upmh.edu.mx,asuarez@fis.cinvestav.mx}
\affiliation{Departamento de Aeron\'autica, Universidad Polit\'ecnica Metropolitana de Hidalgo, Ex-Hacienda San Javier, Tolcayuca, Hgo. C. P. 43860, M\'exico}
\affiliation{Departamento de F\'isica,\\ Centro de Investigaci\'on y
de Estudios Avanzados del IPN, A.P. 14-740, 07000 M\'exico D.F.,
M\'exico.}

\begin{abstract}
In this work the Klein-Gordon (KG) equation for a complex scalar field with $U(1)$ symmetry endowed in a mexican-hat scalar field potential with thermal and electromagnetic contributions is written as a Gross-Pitaevskii (GP)-like equation. This equation is interpreted as a charged generalization of the GP equation at finite temperatures found in previous works. Its hydrodynamical representation is obtained and the corresponding thermodynamical properties are derived and related to measurable quantities. The condensation temperature in the non-relativistic regime associated with the aforementioned system within the semiclassical approximation is calculated. Also, a generalized equation for the conservation of energy for a charged bosonic gas is found when electromagnetic fields are introduced, and it is studied how under certain circumstances its breaking of symmetry can give some insight on the phase transition of the system not just into the condensed phase but also on other related systems. 
\end{abstract}


\maketitle
\section{Introduction}

Since its observation with the help of magnetic traps and lasers (Anderson et al. \cite{Anderson}); the phenomenon of Bose--Einstein condensation has motivated lot of works related to its theoretical and experimental backgrounds. 

In thermodynamics and statistical mechanics, this phenomenon is interpreted as a phase transition where an important matter wave coherence behavior arises from the overlapping of individual de Broglie waves of atoms or particles (bosons), such that a high percentage of these particles condense to the ground state of the system.

In Quantum Field Theory this phenomenon can be related to the spontaneous breaking of a gauge symmetry (Kapusta \cite{Kapusta} and Courteille et al. \cite{Bagnato}). Distiction between broken and unbroken symmetry lines up with the concept of phase transition. In the theory of Bose-condensed systems, there have remained some principal problems that have not been well understood. Many works contain the statement that, though symmetry breaking helps describing BEC's, the latter, in principle, does not require any symmetry breaking. This is however, not correct: Spontaneous breaking of symmetry is a necessary and sufficient condition for Bose-Einstein condensation (Courteille et al. \cite{Bagnato}, Yukalov \cite{Yukalov1}, and Pitaevskii et al. \cite{pitaevskii}).

Symmetry breaking is one of the most essential concepts in particle physics and has been extensively used to study  the behavior of particle interactions in many systems (Pinto et al. \cite{Pinto}). In some cases,  phase transitions  are also identified as changes of states that can be related to changes of symmetries in the system (Griffin et al. \cite{Griffin1}). According to Olivarez-Quiroz, the order of BEC transition in weakly interacting gases predicted by mean-field theory seems to be a first order phase transition, since the relevant  thermodynamic functions do not predict a second-order transition as required by the symmetry-breaking formalism (Olivares-Quiroz et al. \cite{OLI}). Conversely, as it was also suggested and discussed by Lieb (Lieb et al. \cite{lieb}), and references therein (see for example Yukalov \cite{Yukalov1} and Shi et al. \cite{shi}), Bose-Einstein condensation in interacting Bose systems, show spontaneous $U(1)$ gauge symmetry breaking related to second order phase transitions. Thus, is seem that the order of the phase transition associated with Bose-Einstein condensates deserves deeper analysis. 

The study of symmetry breaking mechanisms have turn out to be very helpful in the study of phenomena associated with phase transitions in almost all areas of physics. Bose-Einstein condensation is one topic that uses symmetry breaking mechanisms in an extensive way, and its phase transition associated with the condensation of atoms in the state of lowest energy is the consequence of quantum, statistical and thermodynamical effects. Recently, some results from finite temperature Quantum Field Theory (Dolan et al. and Weinberg \cite{Dolan,Weinberg}) have raised important challenges about the possible physical manifestation of symmetry breaking in condensed matter systems.

Through the analysis of the massive Klein-Gordon equation, the authors (Matos and Castellanos \cite{Matos:2011kn,CastMat}) have explained the way in which a real self-interacting scalar field with a $Z_2$ symmetry can simulate a condensed matter system. They also proved how the Klein-Gordon equation of the scalar field (SF), inside a thermal bath, can be reduced to the Gross-Pitaevskii equation ( which explains the behavior of a Bose-Einstein condensate at zero temperature) in the non-relativistic limit, provided that the temperature of the thermal bath is equal to zero. But the question about the identification of the signature of a condensed system of bosons, with broken symmetry, remains open. The Klein-Gordon equation with a self-interacting one loop scalar field potential also defines a symmetry breaking temperature at which the system may be able to experiment such a phase transition. In this sense, it is ironic how statistical thermodynamics and quantum field theory have not made yet a clear difference between the symmetry breaking mechanism possibly related to Bose-Einstein condensation, and the already known collective behavior of bosonic systems. 

The limited theoretical understanding has made that the relationship between density distributions, phase coherence, and thermal effects on phase transitions become unclear. A deeper study and understanding on how these phenomena can be related is still needed.

To point out some other similarities that can exist between spontaneous symmetry breaking and Bose-Einstein condensation in the case of a field theory with thermal and electromagnetic contributions, is one of the aims of this paper. In this sense, the present work is taken to be complementary to those results reported in (Matos \cite{Matos:2011kn}), (Castellanos \cite{CastMat}), and (Chavanis \cite{chavanis17}). Particularly, this work studies the Klein-Gordon equation of a complex and charged scalar field (with a U(1) symmetry) inside a thermal bath. Again, the hypothesis is that the Klein-Gordon equation, up to one loop in perturbations, might be able to explain the condensation of a scalar field (bosonic system) close to the instant of the phase transition, when the system breaks its $U(1)$ symmetry of the corresponding Lagrangian. 

Finite temperature (and finite density) field theory started out in the 1950's as non relativistic subjects based on quantum mechanics under the name of {\itshape the many-body problem} because it was mainly used in condensed matter and nuclear physics (Fradkin \cite{Fradkin}).

This paper studies a non-homogeneous Bose system. Like in the case of a weakly interacting Bose-Einstein condensate, this work essentially assumes that the particles occupy the same quantum state, and the condensate may be described in terms of a mean-field theory (Andersen \cite{Andersen}). This is in marked contrast to liquid $^4$He, in which a mean-field approach is inapplicable due to the strong correlations induced by the interaction between the atoms. 

Even though most of the cases consider dilute gases, interactions can play an important role as a consequence of the low temperatures, and they give rise to collective phenomena related to Bose-Einstein condensation (see Yukalov \cite{Yukalov}). However, in this work the system treated will not necessarily be weakly interacting and/or diluted, the $\sigma$ term appearing in the equations ahead will account for the contribution of a {\itshape viscosity} that naturally appears through the gradients of the velocity, density, temperature and contributions of external fields. In contrast to an irrotational fluid at zero temperature, it will be seen how the {\itshape viscosity} term appearing in our equations is exactly due to the presence of the thermal bath and external fields. In the case of their absence the classical limit of an irrotational fluid is recovered.

As it will be seen later on, the dynamics of the system are derived directly through Klein-Gordon's equation making the range of validity of the equations ahead much bigger than those obtained through Gross-Pitaevkii's equation, which is obtained mainly for weak interacting gases. 

Finally, the work establishes the conditions and the temperature for the condensation of this non-ideal and thermal Bose gas, coupled to an external field. The calculations seem to show in a direct and simple way some of the essential quantum features of Bose-Einstein condensation.

The paper is organized as follows. In Sec.\ref{sec:SB} a brief theoretical background on the main ideas of gauge symmetry breaking is given, describing how the finite temperature contributions are obtained for the effective potential $V_C$. We then introduce the dynamics of the potential and obtain the new symmetry breaking temperature dependent of the external fields applied. In Sec.\ref{sec:GP} the Gross-Pitaevskii like equation at finite temperatures with electromagnetic contributions is obtained through Klein-Gordon's equation. In Sec.\ref{sec:hydro} the hydrodynamical version of the equations obtained in Sec.\ref{sec:GP} are written. In Sec.\ref{sec:thermodynamic} the thermodynamical equations for the scalar field are derived. In Sec.\ref{sec:TcBEC} the relation between the symmetry breaking temperature and the condensation temperature of the bosons in the non-relativistic limit is calculated, and finally we conclude in Sec.\ref{sec:conclutions}.


\section{Gauge symmetry breaking}\label{sec:SB}

\subsection{Temperature corrections}

In quantum filed theory, many of the times the minimum of the effective potential used to describe the dynamics of a scalar field determines whether symmetry breaking occurs at the quantum level. In his work Coleman (Coleman et al. \cite{Coleman}), started from a classical masseless theory, but found that at the one-loop level, quantum corrections add order $\hbar$ terms to the potential such that the minimum of the effective potential occurs away from the origin.

In the next sections of this work the critical temperature for the phase transition in a model with complex scalar fields and spontaneous symmetry breaking is calculated when the Lagrangian is coupled to electromagnetic fields.

Over time, some approaches to finite temperature field theory have emerged. Some of these are known as: the imaginary-time formalism, the real-time formalism and the thermal field dynamics. In particular, the imaginary-time formalism corresponds to the one which is connected to statistical mechanics. One of the applications of the imaginary time formalism is to show how quantum field theory can give rise to mass corrections proportional to $T^2$ in the potential. A way to quantify this interaction consists in assigning a {\itshape temperature dependent effective mass} to the field, such that this temperature must be of the form $m^2_{Teff}\propto\alpha T^2$ considering dimensional grounds (where $\alpha$ is a proportionality constant of order of unity).

For large T ($m<<T$), it is found that the corrections go like
\begin{equation}
\Delta m^2_{Teff}\propto\lambda(k_BT)^2+O\left(\frac{m}{T}\right).
\end{equation}
In the case of arbitrary $m$, the result is nonanalytic in $m$ and divergent, namely of the form $\Delta m^2_{Teff}=\alpha T^2+\beta\sqrt{m^2}T+\gamma m^2...$ where $\gamma$ diverges. To regulate this inconsistency, dimensional regularization is used, which is needed to keep the dimensions of all terms in the sum equal to that of $\Delta m^2_{Teff}$. When the extra contributions due to non-vanishing temperature are evaluated in the scalar field potential they are found to be finite by themselves, so no problems arise (i.e., the result is a finite sum).  

The $T$-dependence of $\Delta m^2_{Teff}$ in quantum field theory has the physical meaning of the temperature giving extra mass to the scalar particle. The sign leading term ($T^2$ and which does not have any divergence problems) is positive, and is important when studying spontaneously broken fields at finite temperatures. For further analysis  and details on the quantum potential and its corrections up to one loop see (Kolb\footnote[1]{To one loop in quantum corrections, the full potential is given by
$$V_T(\phi_c)=V(\phi_c)+\frac{T^4}{2\pi^2}\int_0^\infty x^2\ln[1-\exp[-(x^2+M^2/T^2)^{1/2}]]$$
where in their notation $V(\phi_c)$ is the zero-temperature one-loop potential (whose contributions are taken into account in the first two terms of Eq. (\ref{eq:V})) and $M$ is a parameter dependent on the mass and the interaction}, Quigg \cite{Kolb,Quigg} and Cervantes \cite{Cervantes}).

\subsection{The potential and the dynamics}

The analysis uses a model containing a local U(1) symmetry with the following Lagrangian,
\begin{eqnarray}
   {\cal L}_\Phi&=&\frac{1}{2}g^{\mu\nu}\left(\nabla_{\mu}\Phi^*-\mathrm{i}\frac{e}{\hbar c}A_\mu\Phi^*\right)\left(\nabla_{\nu}\Phi+\mathrm{i}\frac{e}{\hbar c}A_\nu\Phi\right)\nonumber\\
 &-&V(|\Phi|^2,T)-\frac{1}{4}F^{\mu\nu}F_{\mu\nu}+\mathcal{L}_{int}\,,
  \label{eq:L}
  \end{eqnarray}
where $F_{\mu\nu}$ is the Maxwell tensor. The metric signature used is $(+,-,-,-)$. The simplest case of a double-well Mexican-hat potential up to one-loop contributions for a complex self-interacting scalar field, $\Phi(\boldsymbol{r},t)$ (see Higgs \cite{Higgs}) defines a scalar field potential $V_c$ as follows
\begin{eqnarray}
V_C(|\Phi|^2,T)=&-&\frac{m^2c^2}{2\hbar^2}|\Phi|^2+\frac{\lambda}{4\hbar c}(|\Phi|^2)^2+\nonumber\\
&+&\frac{\lambda}{8\hbar^2c^2}(k_BT)^2|\Phi|^2-\frac{\pi^2}{90\hbar^3c^3}(k_BT)^4,\nonumber\\
\label{eq:V}
\end{eqnarray}
where $V_C$ will stand for the standard one-loop potential with temperature contributions, see for example (Calzetta \cite{Calzetta}) and others (Gaberdiel \cite{Gaberdiel}, Kolb \cite{Kolb})

In Eq.(\ref{eq:V}), $m$ represents the mass parameter related to the true physical mass of the the boson, which will be determined by the curvature of the potential once the field breaks its symmetry and its new stable ground state is found (possibly in the condensed phase, see for example Kolb \cite{Kolb}, and Su\'arez \cite{suarez3}), $\lambda$ describes the strength of the interaction, $k_B$ is the Boltzmann's constant, $\hbar$ is Planck's constant, $T$ is the temperature of the thermal bath and from hereafter the speed of light, $c$, will be equal to one in the case of not being explicitly written.

As mentioned previously, the phenomenon of symmetry breaking can be understood in several ways. If one naively attempts to construct a theory using only the potential in Eq.(\ref{eq:V}) without considering the temperature contributions, then one finds that the negative quadratic term of the mass in the scalar field becomes the most important contribution to the potential. However, if the field $\Phi$ is supposed to be in contact with a thermal bath, the interaction of the particles (bosons) with the thermal bath can, in general, counteract this term. Then, at finite temperatures the effective mass of the scalar field turns out to be positive, as long as the effective mass due to the thermal bath is greater than the mass due just to the bosonic particles.

Note that the potential is defined by the the sum of the classical expression ($V_{class}\sim m^2\Phi\Phi^*+\lambda(\Phi\Phi^*)^2$) plus contributions produced by quantum effects. Correction terms tend to destabilize the symmetry. Therefore, quantum corrections are said to spontaneously produce the breaking of symmetry. Quantum corrections can then influence the structure of the potential, and hence the dynamics of the system (in our case, the temperature is not the only information for the phase transition, but also the external fields $\boldsymbol{A}$ and $\Phi$, which can also make a contribution).

The situation is rather different for temperatures below a symmetry breaking critical temperature, $T_c^{SB}$, when the system is supposed to be in the condensed phase. As the temperature drops below the temperature $T_c^{SB}$, due to cooling or applied external fields, then the mass term with negative sign dominates, and the system develops a new minimum at very low temperatures ($T<<T_c^{SB}$), which can lead to spontaneous symmetry breaking (SSB).  The change of temperature due to cooling or applied external fields can produce a privileged phase, which will tend to rearrange the system.

Consequently, the physical meaning of SSB can be seen as follows: for large temperatures, $T>>T_c^{SB}$, finite temperature quantum corrections maintain the field located in its minimum $\Phi=0$. There, the system contains symmetries which are reflected on the invariant properties of the Lagrangian (in this case $\Phi\rightarrow-\Phi$). Due to some process (application of external fields for example) the system cools down until reaching some critical temperature, where the quantum contributions of the potential start to be important, and the potential develops a new minimum with $\Phi\neq0$. The $\Phi$ field tends tends to roll down the hill of its potential, towards this new minimum (possibly associated with the condensed phase). During the symmetry breaking the field $\Phi$ acquires different values and evolves to the more energetically favorable state; particles lose their symmetric state and start to interact strongly between them, and this is the point where it is said that the condensation has occurred. 

Notice that Eq.(\ref{eq:L}) contains a term with a Lagrangian of interaction $\mathcal{L}_{int}$ such that $\mathcal{L}_{int}=\frac{m^2}{\hbar^2}\phi|\Phi|^2$. A first-order interaction  potential $\phi$ is introduced (i.e, a scalar potential that does not depend on powers of $\phi$) ; this potential will represent the external trapping potential for the bosonic system (an external magnetic field or a laser for example, see Su\'arez et al. \cite{Matos:2011pd}).
 
For a charged field the D'Alambertian operator is given by  $\Box_E^{2}\equiv\left(\nabla_{\mu}+\mathrm{i}e/\hbar\ A_\mu\right)\left(\nabla^{\mu}+\mathrm{i}e/\hbar\ A^\mu\right)\label{eq:Box}$, where $A_\mu=(\boldsymbol{A},\varphi)$ is the electromagnetic four-potential. It is convenient to define a total potential $V_T$, which will add the external potential $\phi$ to the potential of the SF,
\begin{equation}
V_T(|\Phi|^2,T)=V_C(|\Phi|^2,T)
-\frac{m^2}{\hbar^2}\phi{|\Phi|^2}.
\label{eq:VT}
\end{equation}

Thus, from Eq.(\ref{eq:VT}), an effective mass for the $\Phi\Phi^*=|\Phi|^2$ term of the scalar field at $T=0$ can also be defined,
\begin{equation}
V_T(|\Phi|^2,T=0)=-\frac{m_{eff}^2}{\hbar^2}|\Phi|^2+\frac{\lambda}{2\hbar^2}(|\Phi|^2)^2
 \end{equation}
 such that
 \begin{equation}
 m_{eff}=\sqrt{m^2(1+2\phi)}.
 \label{eq:KGT}
 \end{equation}
 and $V_T$ can be rewritten as
 \begin{eqnarray}
 V_T(|\Phi|^2,T)=&-&\frac{m_{eff}^2}{\hbar^2}|\Phi|^2+\frac{\lambda}{2\hbar^2}(|\Phi|^2)^2\nonumber\\
&+&\frac{\lambda}{4\hbar^2}(k_BT)^2|\Phi|^2-\frac{\pi^2}{90\hbar^3}(k_BT)^4
 \label{eq:potT}
 \end{eqnarray}

With this equation at hand, for the $V_T$ potential the critical temperature of symmetry breaking, $T^{SB}_{c}$,  where the minimum of the potential at $\Phi=0$ becomes a maximum and at which the symmetry is found to be broken, will now be given by the following expression
 \begin{equation}
  k_B T^{SB}_{c}=\frac{2}{\sqrt{\lambda}}\sqrt{m_{eff}^2-\frac{\lambda}{\kappa^2} n}, 
\label{eq:Tc}
 \end{equation}
where $\kappa^{-2}n=\Phi\Phi^*=|\Phi|^2$ (related to the rest-mass density of the system, see Su\'arez \cite{suarez2}). In equation (\ref{eq:Tc}) the field $\Phi$ and consequently $|\Phi|^2=n$ cannot be strictly equal to zero, since the field slowly oscillates around the new minimum of the potential with small amplitudes (although it would seem that the temperature is only well defined when $n\approx0$ or constant). Additionally, $T^{SB}_{c}$ is only well defined locally, for instance, at the center of the trapping potential (Castellanos \cite{CastMat}). In other words, $n$ can be set equal to zero as an approximation, and consequently the symmetry breaking temperature can be written as $ T^{SB}_{c}\approx \frac{2 m_{eff}}{k_{B} \sqrt{\lambda}}$. However, strictly speaking, the contributions of $n$ must be taken into account since the field is not exactly zero at the minimum of the scalar field's potential after the transition (Kolb \cite{Kolb}). 
 Moreover, the transition defined by Eq.\,(\ref{eq:Tc}) has to be carried out through a nearly quasi--static process, so that the temperature of symmetry breaking can be well defined only locally, for instance, near the center of the trap as was mentioned above.
 
We are also assuming slow time-varying fields, very close to the static case where we are in local equilibrium, i.e., for specific conditions upon the density $n$ and the field $\phi$. In fact, the electromagnetic field and temperature are constant, they only vary very slowly conserving the system in quasi-thermal equilibrium. In this case there is no need of gauge fixing, and an specific problem should determine the convenient gauge to work with.

It is well known that, for these kind of trapped systems, the bosons can be packed so closely together that the system results strong interacting once the condensed phase has been reached (Yukalov \cite{Yukalov}). Eq.(\ref{eq:Tc}) shows how the breaking of symmetry exhibits the same behavior as the condensed phase, since if the symmetry breaking temperature $T^{SB}_{c}$ is high enough, then the interaction parameter $\lambda$ turns out small, i.e. the condensed phase has not been reached and the bosons are not gathered in the same state, being distributed also along the excited states of the system. On the other hand, if the temperature of symmetry breaking is small (possibly equal to the temperature of condensation), then this situation is usually related to the fact that the interaction becomes large and most of the bosons occupy the same quantum state. In this case, it can now be seen how these interactions can also be controlled through the manipulation of the fields $A$ a $\phi$.

A Bose Einstein Condensate (BEC) can also be considered as consisting of charged atoms, that is, of ions which experience electromagnetical self-interaction. Some experiments have been carried out studying these kind of systems, for example, it has been seen how charged rotating BECs are energetically more favorable for forming vortices, it is also interesting for the study of superconductivity, etc. (Kamura et al.\cite{Kumakura}, and Balewski et al. \cite{Balewski})

At this point, it is important to mention some issues related to the Meissner effect. It is true that if we assume  a charged Bose gas system with short-range repulsive interactions then, the Meissner effect must be present below the temperature $T_{c}^{SB}$. In the case of complex SF's, if its charge density is neutral, the bosons do not experience any Meissner effect (electromagnetic rotations) even when a magnetic field is present. In the case the charge density is not neutral, it is exactly the mass acquired through the process of symmetry breaking that can possibly give rise to Meissner effects . However, a detailed analysis concerning the Meissner effect for our system will be done in a forthcoming work. 
It has also been argued that because of particle interactions these kind of systems can become a superfluid (see for instance Stwalley \cite{Stwalley}). Significant progress has been made in the last twenty years, but there are still many open problems to be solved on the subject.


\section{The generalized Gross-Pitaevskii equation}\label{sec:GP}

The Klein-Gordon (KG) equation without self-interactions (no electromagnetic fields, and at $T=0$) can be viewed as a relativistic generalization of the Schr\"odinger equation. Similarly, the KG equation with self-interactions can be viewed as a relativistic generalization of the Gross-Pitaevskii equation (GP).  In order to recover the Schr\"odinger and GP equations in the nonrelativistic limit of the formalism, the following \emph{ansatz} is made
 \begin{equation}
\kappa\Phi({\bf r},t)=\Psi({\bf r},t)\exp^{-{\mathrm i}mc^2t/\hbar},
\label{eq:Phi}
 \end{equation}
where $\kappa$ is a normalization parameter related to the mass of the bosons (the justification of such prefactor can be found in Su\'arez \cite{suarez2} and Zee \cite{Zee}). In this section we generalize our previous results even further by introducing electromagnetic field contributions to the SF theory (KG equation) at non-zero temperatures.

We can always use the Klein transformation (Eq.(\ref{eq:Phi})) when calculating the terms $\partial_\nu\Phi$ appearing in the KG equation of the SF coupled to electromagnetic fields at zero temperature. In this case the following identity for the covariant derivatives of $\Phi$ in the KG equation is used,
$$D_\mu\partial_\nu\Phi=\partial_\mu\partial_\nu\Phi-\Gamma_{\mu\nu}^\sigma\partial_\sigma\Phi$$
(see \cite{chavanis17} for a detailed derivation of the electromagnetic KG equation at zero temperature). Then, in terms of the complex function $\Psi$ and the total potential $V_T$ (Eq.(\ref{eq:potT})), the Klein-Gordon equation in ordinary units for this case reads as follows,
\begin{eqnarray}
 \mathrm{i}\hbar\dot{\Psi}&+&\frac{\hbar^2}{2m}\Box_E^2{\Psi}+\frac{1}{2}mc^2\Psi-eA_0\Psi\nonumber\\
&-&\frac{m^2_{eff}}{2m}\Psi+
\frac{\lambda n}{2m\kappa^2}\Psi+\frac{\lambda}{8m}(k_BT)^2\Psi=0.
\label{eq:GP}
\end{eqnarray}
Here we make the assumption that the relationship $|\Psi|^2=\kappa^2\Phi\Phi^*=n$ is satisfied. Equation (\ref{eq:GP}) is exact and describes the field $\Psi(\boldsymbol{r},t)$ at finite temperatures in interaction with electromagnetic fields. 

In order to interpret Eq.(\ref{eq:GP}) as a generalization of the Gross-Pitaevskii equation, notice also that a relation between the interaction parameter $\lambda$ and the s-wave scattering length $a_{s}$ is needed. The scalar field, $\Psi(\boldsymbol{r},t)$, plays the role of an order parameter, as in the usual Gross--Pitaevskii equation. For the Gross-Pitaevskii equation in the mean field limit (s-wave scattering length much smaller to the average distance between particles), the interaction parameter, usually called $g$ is related to the s-wave scattering length through $g=4\pi\hbar^2a_s/m$ at the lowest Born approximation. With this evidence at hand, by comparison with Eq.(\ref{eq:GP}) we have
$$g\propto\frac{\lambda}{2m\kappa^2}$$
such that 
$$\lambda=8\pi\hbar^{2}a_s\kappa^{2}.$$ 
The critical temperature of symmetry breaking (Eq.(\ref{eq:Tc})) can then be directly related to the s-wave scattering length $a_s$ through this relation (Castellanos {\itshape et al}. and Chavanis \cite{CastMat,chavanis}). 

To obtain a measurement of the corresponding symmetry breaking temperature in classical systems, close to the critical temperature for Bose-Einstein condensation, which should be very small or near to zero, these facts suggest that dense systems, together with small bosonic masses are needed. Furthermore, since $T_c^{SB}$ also depends on $m_{eff}$ then these results can probably be combined in some way in order for $T_c^{SB}$ to be small, even for a weak coupling constant and large masses. In this case (according to Eq.(\ref{eq:KGT}) and Eq.(\ref{eq:Tc})), $\lambda n/\kappa^2$ has to take at most the value of $m_{eff}$ giving the following restriction for the strength of the field of the trap
\begin{equation}
\phi\geq\left(\frac{\lambda}{m^2\kappa^2}\right)n.
\label{eq:TR}
\end{equation}

Equation (\ref{eq:TR}) tells us explicitly that the strength of the field applied in the trap in order to generate a BEC will depend on the mass of the particles used (in this case bosons) and their scattering length, i. e., the characteristics of the trapping field $\phi$ depend on the type of particles that are being considered for condensation. As an example the frequency of a harmonic trap depends upon the mass parameter of the particles. Three different groups have reported the direct evidence of BEC in weakly interacting systems of atoms such as rubidium (Anderson et al. \cite{Anderson}), lithium (Bradley et al. \cite{bradley}), and sodium (Davis et al. \cite{davis}), confined in harmonic traps and cooled down to very low temperatures. They show that the main effect of interactions is to reduce the density of the cloud of particles. These experiments have experimentally studied some of the aspects of the theory, such as symmetry breaking. 

\section{The Hydrodynamical version}\label{sec:hydro}

Next, Eq.(\ref{eq:GP}) is transformed into its hydrodynamical version (Chiueh and Bohm \cite{Chiueh,Bohm}). For this purpose, the function $\Psi$ will be represented in terms of a modulus $n$ and a phase $S$ as follows,
 \begin{equation}
 \Psi=\sqrt{ n}\,\mathrm{exp}({\mathrm{i}S/\hbar}),
 \label{eq:psi}
 \end{equation}
where the phase $S(\boldsymbol{r},t)$ is defined as a real function. 

Here $n(\boldsymbol{r},t)=\rho/M_T$ is interpreted as the number density of particles in the condensed plus thermal states (excited states), being $M_T $ the total mass of the particles in the system. Both, $S$ and $n$ are functions of time and position. 

Below the critical temperature $T^{SB}_{c}$, the density will oscillate around $\rho=\kappa^{-2} k_B^2((T^{SB}_{c})^2-T^2)/4$, which can be different from zero as long as $T\neq T^{SB}_{c}$ (Matos \cite{Matos:2011kn} and Su\'arez\cite{suarez3}). Below this transition, a two component system is expected, with a dense central region possibly surrounded by a diffuse, non-condensed fraction. With the application of Madelung's transformation (Eq.\eqref{eq:psi}) for Eq.\eqref{eq:GP}, and after separating the real and imaginary parts we obtain
 \begin{subequations}
 \begin{eqnarray}
 \dot{ n}&+&\boldsymbol{\nabla}\cdot\left[n\frac{\hbar}{m}\left(\boldsymbol{\nabla S}-\frac{e}{\hbar}\boldsymbol{A}\right)\right]-\left[n\frac{\hbar}{mc}\left(\dot S-\frac{e}{\hbar}\varphi\right)\right]^{\cdot}=0,\nonumber\\
 \label{eq:ntotal}
  \end{eqnarray}
   \begin{eqnarray}
&& \frac{\hbar}{m}\left(\dot S-\frac{e}{\hbar}\varphi\right)+\frac{\lambda}{2m^2\kappa^2} n+(\phi+1) +\frac{\lambda}{8m^2}k_B^2T^2\nonumber\\
&+&\frac{1}{2}\left[\left(\frac{\hbar}{m}\left(\boldsymbol{\nabla}S-\frac{e}{\hbar}\boldsymbol{A}\right)\right)^2-c\left(\frac{\hbar}{m c}\left(\dot S-\frac{e}{\hbar}\varphi\right)\right)^2\right]\nonumber\\
&-&\frac{\hbar^2}{2m^2}\left(\frac{\Box^2\sqrt{ n}}{\sqrt{ n}}\right)=0.
   \label{eq:Stotal}
 \end{eqnarray}\label{eq:hidro}
  \end{subequations}
If we apply the gradient operator to Eq.(\ref{eq:Stotal}) and use the following definitions for the scalar flux (not a vector) and the velocity field, respectively
\begin{equation}
j=n\frac{\hbar}{m c}\left(\dot S-\frac{e}{\hbar}\varphi\right)
\label{eq:fluxesj}
\end{equation}
 \begin{equation}
  \boldsymbol{v}\equiv\frac{\hbar}{m}\left(\boldsymbol{\nabla}S-\frac{e}{\hbar}\boldsymbol{A}\right),
 \label{eq:vel}
 \end{equation}
then the set of Eqs. in \eqref{eq:hidro} can be written as follows,
\begin{subequations}
 \begin{eqnarray}
  \dot{ n}+\boldsymbol{\nabla}\cdot( n\boldsymbol{v})
  -\dot j&=&0,\label{eq:cont}
\end{eqnarray}
\begin{eqnarray}  
  \dot{\boldsymbol{v}}+(\boldsymbol{v}\cdot\boldsymbol{\nabla})\boldsymbol{v}&=&\frac{e}{m}(\boldsymbol{E}+\boldsymbol{v}\times\boldsymbol{B})-\boldsymbol{\nabla}\phi-\frac{\lambda}{2m^2\kappa^2}\boldsymbol{\nabla}n\nonumber\\
&+&\frac{\hbar^2}{2m^2}\left[\boldsymbol{\nabla}\left(\frac{\nabla^2\sqrt{ n}}{\sqrt{ n}}\right)-\frac{1}{c^2}\boldsymbol{\nabla}\left(\frac{\partial^2_t\sqrt{ n}}{\sqrt{ n}}\right)\right]\nonumber\\
&-&\frac{\lambda k^2_B}{4m^2}T\boldsymbol{\nabla}T+\frac{\hbar^2}{2m^2c}\boldsymbol{\nabla}\left(\dot S-\frac{e}{\hbar}\varphi\right)^2,\nonumber\\
  \label{eq:2}
 \end{eqnarray}\label{eq:hydro}
\end{subequations}
where $\boldsymbol{E}=-\partial\boldsymbol{A}/\partial t-\nabla\cdot\varphi$, and $\boldsymbol{B}=\nabla\times\boldsymbol{A}$ are the electric and magnetic field vectors, respectively. The constant $\hbar$ enters on the right-hand side of Eq. (\ref{eq:2}) through the third and last terms, which now not only contain factors accounting for the quantum potential, but they also contain factors related to the electromagnetic field and to the relativistic contributions. Also, from (15) it shown that when electromagnetic fields are present, the fluid in question is not longer irrotational since the relation $\boldsymbol\nabla\times\boldsymbol v=0$ for the fluid is not longer true.

If we now multiply Eq.(\ref{eq:2}) by $ n$, then
 \begin{eqnarray}
 n\dot{\boldsymbol{v}}+n(\boldsymbol{v}\cdot\boldsymbol{\nabla})\boldsymbol{v}=n\boldsymbol{F}_E+n\boldsymbol{F}_\phi -\boldsymbol{\nabla}p
  +n\boldsymbol{F}_Q+\boldsymbol{\nabla}\sigma,\nonumber\\
 \label{eq:navier}
 \end{eqnarray}
where $\boldsymbol{F}_E$ can be identified with the electromagnetic force, $\boldsymbol{F}_\phi=-\boldsymbol{\nabla}\phi$ is the force associated with the external potential $\phi$, $p$ can be seen as the pressure of the scalar field gas satisfying the equation of state $p=\omega n^2$,
\begin{equation}
\boldsymbol\nabla p=\boldsymbol\nabla\left(\frac{\lambda}{4m^2\kappa^2}n^2\right)=\frac{n\lambda}{2m^2\kappa^2}\boldsymbol\nabla n,
\end{equation}
$\boldsymbol{\nabla}p$ are forces produced by the gradients of pressure, where $\omega=\lambda/(4m^2\kappa^2)$, and $\boldsymbol{F}_Q=-\boldsymbol\nabla U_Q$ is the quantum force associated with the quantum potential (Grossing and Pethick et al. \cite{gro,Pethick}),
\begin{equation}
\boldsymbol F_Q=-\boldsymbol\nabla U_Q=\frac{\hbar^2}{2m^2}\boldsymbol\nabla\left(\frac{\nabla^2\sqrt n}{\sqrt n}\right).
\end{equation}
 
Finally, $\boldsymbol{\nabla}\sigma$ is defined as follows
\begin{eqnarray}
 \boldsymbol{\nabla}\sigma&=&\frac{\hbar^2}{2m^2c}n\boldsymbol{\nabla}\left(\dot S-e\varphi\right)^2
 -\frac{1}{4}\frac{\lambda}{m^2}k_B^2 nT\boldsymbol{\nabla}T\nonumber\\
 &+&\zeta\boldsymbol{\nabla}(\ln n{\dot)}-\frac{\hbar^2 n}{4m^2c^2}\boldsymbol{\nabla}\left(\frac{\ddot{n}}{n}\right),
 \label{eq:sigma}
 \end{eqnarray}
where the coefficient $\zeta$ is now given by the following equation
\begin{equation}
 \zeta=\frac{\hbar^2}{4m^2}\left[-\boldsymbol\nabla\cdot( n\boldsymbol{v})+\dot{j}\right].
 \label{eq:zeta}
 \end{equation}

Therefore, from Eq.(\ref{eq:sigma}), the total energy of the system will have an additional contribution that will come from a charged flux viscosity $\nabla\sigma$  (Matos et al. \cite{Matos:2011kn}); as it will be seen later on in Section \ref{sec:thermodynamic}.

The non-relativistic hydrodynamics based on the Schr\"odinger or Gross-Pitaevskii equations are recovered in the limit $c\rightarrow\infty$, which is exactly the non-relativistic limit of the theory. Inside this limit, the system of equations in (\ref{eq:hydro}) is given by
\begin{subequations}
 \begin{eqnarray}
  \dot{ n}+\boldsymbol{\nabla}\cdot( n\boldsymbol{v})&=&0,\label{eq:contNR}\\
  n\dot{\boldsymbol{v}}+n(\boldsymbol{v}\cdot\boldsymbol{\nabla})\boldsymbol{v}&=&n\boldsymbol{F}_E+n\boldsymbol{F}_\phi
  -\boldsymbol{\nabla}p+n\boldsymbol{F}_Q+\boldsymbol{\nabla}\sigma,\nonumber\\
   \label{eq:navierNR}
 \end{eqnarray}\label{eq:hydroNR}
 \end{subequations} 
see also Eq.(\ref{eq:fluxesj}). Here Eq.(\ref{eq:contNR}) is the continuity equation, and Eq.(\ref{eq:navierNR}) is the equation for the momentum (Su\'arez et al. \cite{Matos:2011pd}). Eq.(\ref{eq:navierNR}) which describes the vector ${\boldsymbol v}({\boldsymbol r},t)$, is the velocity of the flow, which as mentioned before, turns out to be irrotational when $T=0$ and there are no electromagnetic fields. It is also worth pointing out that the equation of continuity and the equation for the velocity provide a set of coupled equations, exactly equivalent to the generalized GP equation given by Eq. (\ref{eq:GP}).

From equations (\ref{eq:sigma}) and (\ref{eq:zeta}), it is possible to see that in the classical limit of the Klein-Gordon equation, at temperature $T=0$ and nonexistent electromagnetic fields, the flux viscosity is given by the following equation 
 \begin{equation}
 \boldsymbol{\nabla}\sigma=-\left[\frac{\hbar^2}{4m^2}\boldsymbol\nabla\cdot(n\boldsymbol{v})\right]\boldsymbol\nabla(\mbox{ln }n)\dot{},
 \end{equation}
which is directly related  to the gradient of the velocity, and this is just the definition of a flux viscosity in non ideal fluids.

Eqs.(\ref{eq:hydroNR}) show the presence of an anisotropic  and non-thermal velocity distribution at $T=0$, which is expected for the minimum quantum state of energy once $T<<T_{c}^{SB}$. In contrast, a thermal velocity distribution is obtained for $T\neq 0$, i.e, even when there might exist a breaking of symmetry in the system, excited states may be identified within the condensed state. From Eqs. (\ref{eq:hydroNR}) we find there is an interesting parallel between symmetry breaking and Bose-Einstein condensation. 

In usual laboratory Bose--Einstein condensates, the Gross-Pitaevskii equation is an approximate equation, which
describes the properties of the bosonic system at temperatures much smaller that the condensation
temperature (strictly speaking at zero-temperature), valid also when the scattering length $a_s$ (related
to our self-interaction parameter $\lambda$) is much smaller than the mean inter-particle spacing, that is,
the gas is sufficiently diluted. In other words, there are not temperature corrections when we assume
the validity in describing the condensate with the Gross-Pitaevskii equation. In this sense, the
contribution of the external bath on the effective Gross-Pitaevskii Eq. (\ref{eq:GP}) deduced from the Klein-Gordon equation, can be interpreted as a finite temperature correction for the system,
which might not be in equilibrium with the thermal bath. This is the same behavior as in usual condensates, in
which we have the condensed phase immersed in a cloud of particles which are situated out of the
ground state, i,e., there is always a thermal cloud (depletion), that under typical experimental conditions
could be neglected, (see Pethick et al. \cite{Pethick}). Moreover, notice that in the non-relativistic limit, the
d'Alambertian operator in Eq. (\ref{eq:GP}) becomes the Laplacian operator and setting T = 0 (with no electromagnetic fields), Eq. (\ref{eq:GP}) becomes the usual Gross-Pitaevskii equation describing Bose-Einstein condensation.


\section{The Thermodynamics}\label{sec:thermodynamic}

In what follows, some of the thermodynamical equations that represent the previous system will be derived through their hydrodynamical representation. To do this, we use the conservation equation for the quantum potential $U_Q$ satisfying the following relation,
  \begin{equation}
    (n U_Q)\dot{}+\boldsymbol{\nabla}\cdot (n U_Q\boldsymbol{v}+\boldsymbol{J}_\rho)+ n\boldsymbol{v}\cdot\boldsymbol{F}_Q=0
   \label{eq:contUQ21}
  \end{equation}
which follows by direct calculation (see \cite{Matos:2011kn} for more details). Eq.(\ref{eq:contUQ21}) uses the quantum density flux $\boldsymbol{J}_\rho= n\boldsymbol{v}_\rho\nonumber$,  where the term $\boldsymbol{v}_\rho=\frac{\hbar^2}{4m^2}({\boldsymbol\nabla}\mbox{ln}n)\dot{}$ defines a velocity field related to the number density of particles. In (Matos et al. \cite{Matos:2011kn}) the authors  interpret this velocity as a flux produced only by the potential $U_Q$.

The total energy of the system will be the sum of the energies of each of the contributions, where the total energy density of the system $\epsilon$ is the sum of the kinetic, potential and internal energies (Oliver et al. \cite{oli}), in this case we have an extra term $U_Q$ due to the quantum potential
 \begin{equation}
  \epsilon=\frac{1}{2} nv^2+ n\phi+ nu+ nU_Q+\psi_E
 \label{eq:energiae}
 \end{equation}
being $u$ the inner energy of the system and 
 \begin{equation}
  \psi_E=\frac{e}{m}(\varphi-\boldsymbol{v}\cdot\boldsymbol{A})
 \label{eq:psiE}
 \end{equation}
the electromagnetic energy potential, defined in terms of the vector potential $\boldsymbol{A}$ and the electric potential $\varphi$.

To obtain the continuity equation for $\boldsymbol{A}$ observe that
\begin{eqnarray}
  \boldsymbol{F}_E&=&\frac{e}{m}\left(-\frac{\partial\boldsymbol{A}}{\partial t}-\boldsymbol{\nabla}\varphi+\boldsymbol{v}\times\boldsymbol{\nabla}\times\boldsymbol{A}\right)\nonumber\\
  &=&\frac{e}{m}\left(-\boldsymbol{\nabla}\varphi+\boldsymbol{\nabla}(\boldsymbol{v}\cdot\boldsymbol{A})\right.\nonumber\\
  &-&\left.\left[\frac{\partial\boldsymbol{A}}{\partial t}+(\boldsymbol{v}\cdot\boldsymbol{\nabla})\boldsymbol{A}+(\boldsymbol{A}\cdot\boldsymbol{\nabla})\boldsymbol{v}\right]\right)=-\boldsymbol{\nabla}\psi_E-\boldsymbol{j}_B.\nonumber\\
  \label{tonaa}
\end{eqnarray}
By using Eq.(\ref{tonaa}), we get the following result
 \begin{eqnarray}
  (n\psi_E\dot)&=&\dot n\psi_E+n\dot\psi_E\nonumber\\
  &=&-\boldsymbol{\nabla}\cdot(n\boldsymbol{v}\psi_E)+n\boldsymbol{v}\cdot\boldsymbol{\nabla}\psi_E+n\dot\psi_E\nonumber\\
  &=&-\boldsymbol{\nabla}\cdot(n\boldsymbol{v}\psi_E)-n\boldsymbol{v}\cdot\boldsymbol{F}_E-n\boldsymbol{v}\cdot\boldsymbol{j}_B+n\dot\psi_E\nonumber
  \end{eqnarray}
where we have used the continuity equation for $n$. 
Thus $\psi_E$ fulfills the continuity equation 
\begin{equation}
(n\psi_E\dot)+\boldsymbol{\nabla}\cdot(n\boldsymbol{v}\psi_E+\boldsymbol{J}_B)=n\dot\psi_E-n\boldsymbol{v}\cdot \boldsymbol{F}_E,
 \label{eq:psiE}
 \end{equation}
such that $\nabla\cdot\boldsymbol{J}_B=n\boldsymbol{v}\cdot\boldsymbol{j}_B$ is given by the continuity equation of the vector potential $\boldsymbol{A}$.
\begin{equation}
 \frac{\partial \boldsymbol{A}}{\partial t}+(\boldsymbol{v}\cdot\boldsymbol{\nabla}) \boldsymbol{A}=-(\boldsymbol{A}\cdot\boldsymbol{\nabla})\boldsymbol{v}+\frac{m}{e}\boldsymbol{j}_B,
   \label{eq:contA}
  \end{equation}
From Eq.(\ref{eq:energiae}) the internal energy $u$ will then satisfy the equation
  \begin{equation}
     \left(n u\right)\dot{}+\boldsymbol{\nabla}\cdot(n\boldsymbol{v}u+\boldsymbol{J}_q+\boldsymbol{J}_B-p\boldsymbol{v}-\boldsymbol{J}_\rho)=
   -p\boldsymbol{\nabla}\cdot\boldsymbol{v},
    \label{eq:conte}
  \end{equation}
such that, 
$\boldsymbol{\nabla}\cdot\boldsymbol{J}_q=\boldsymbol{v}\cdot(\boldsymbol{\nabla}\sigma)$
and
$\boldsymbol{\nabla}\cdot\boldsymbol{J}_B=\boldsymbol{v}\cdot(n\boldsymbol{j}_B)$.

In order to find the thermodynamical quantities of the system in equilibrium (taking $p$ as constant on a volume $L$), the system is constrained to a regime where the trapping potential is constant in time.

The integration of Eq.(\ref{eq:conte}) on a closed region provides the following result,
  \begin{eqnarray}
  \frac{\mathrm{d}}{\mathrm{d}t}\int n u\,\mathrm{d}V&+&\oint (\boldsymbol{J}_q+\boldsymbol{J}_B+p\boldsymbol{v})\cdot\boldsymbol{n}\, \mathrm{d}S-\oint\,\boldsymbol{J}_\rho\cdot\boldsymbol{n}\, \mathrm{d}S\nonumber\\
   &=&-p\frac{\mathrm{d}}{\mathrm{d}t}\int \,\mathrm{d}V.
   \label{eq:contu2}
  \end{eqnarray}
This equation is the reason that the expression describing the conservation of energy of the Klein-Gordon equation reads as follows,
  \begin{equation}
   \mathrm{d}U=\text{\^d} Q+\text{\^d} Q_B+\text{\^d}A_Q-p\mathrm{d}V
   \label{eq:1leyBEC}
  \end{equation}
where $U=\int n u\,\mathrm{d}V$ is the internal energy of the system, (Pitaevskii et al. \cite{b.n}); and as it can be seen (Eq.(\ref{eq:contu2})), its change is the result of a combination of the heat $Q$ added to the system and work made on the system (pressure dependent).

Making use of the divergence theorem of vector calculus and Eq.(\ref{eq:contUQ21}), we have
$$\frac{\hat{d}A_Q}{dt}=\int\nabla\cdot(n\boldsymbol{v}_\rho)dV=-\int n\dot{U}_QdV,$$
so that
$$\hat dA_Q=-ndU_Q.$$
A careful observation of the last equation suggests that the quantities related to the quantum potential (and number density) are point functions of the thermodynamical system, i.e., their value depend on the path on which the equilibrium state can be reached, but not on the initial and final states of the system (Matos \cite{Matos:2011kn}). 

Therefore, quantum corrections effects related to the phase transition of Bose-Einstein condensation seem to be an intrinsic property which does not depend on the experiment. Also, from Eq.(\ref{eq:1leyBEC}) the change in energy seems directly affected by the temperature of the system or viceversa.   

Analogously, for the magnetic contribution, we obtain the following equation,
\begin{eqnarray}
   \frac{\text{\^d}Q_B}{\mathrm{d}t}&=&\int \boldsymbol{\nabla}\cdot\boldsymbol{J}_B\,dV= \int\boldsymbol{v}\cdot(n\boldsymbol{j}_B)\,dV\nonumber\\
  &=&\frac{m}{e}\int  n\left[\frac{\partial \boldsymbol{A}}{\partial t}+(\boldsymbol{v}\cdot\boldsymbol{\nabla}) \boldsymbol{A}+(\boldsymbol{A}\cdot\boldsymbol{\nabla})\boldsymbol{v}\right]\cdot\boldsymbol{v}\,dV, \nonumber\\
   \label{eq:dQ_B}
  \end{eqnarray}
where the vector potential $\boldsymbol{A}$ fulfills the Maxwell equations. Observe how the fluxes contain information of the velocity of the fluid and of the electromagnetic contributions, this point might be important in the study of superconductivity for example.

Superfluidity is a phenomenon also strongly related to Bose-Einstein condensation. As mentioned before, from the definition in Eq.(\ref{eq:vel}), it can be seen that the velocity field of the scalar field fluid then results irrotational in the case of non-existent external electromagnetic fields. If dissipative processes such as viscosity (which is related to $\sigma$) and thermoconductivity (which is related to $T\nabla T$) are absent, then the system assumes a superfluid like behavior.

When playing with the conditions of the system, we might be able to find situations with similar results to those found in superconductivity or superfluidity. Clearly, there is a need for more theoretical calculations in the transition regimes to understand in a better way the relationship between coherence, BEC's, superconductivity and superfluidity.  


\section{The condensation Temperature (Non--Relativistic)}\label{sec:TcBEC}

This section contains the computation of the condensation temperature $T_c$ (which must not be mistaken with the temperature of symmetry breaking $ T^{SB}_{c}$) in the non-relativistic regime associated with the aforementioned system, within the semiclassical approximation is calculated (Pethick, Pitaevskii and Dalfovo et al. \cite{Dalfovo,b.n,Pethick}). 

The analysis of a Bose-Einstein condensates in the ideal case
and with a finite number of particles, trapped in different potentials (Bagnato, Giorgini and Grossmann et al. \cite{bagnato,grossmann,Giorgini} and references therein) shows that the main properties associated with the condensate, and in particular the condensation temperature, depend strongly on the characteristics of the trapping potential in question, the number of spatial dimensions, and the functional form of the corresponding
single-particle energy spectrum. 

Inserting plane waves in the corresponding Klein-Gordon equation, and neglecting the term proportional to $T^{4}$ in Eq.(\ref{eq:V}) (assuming that the temperature is sufficiently small), plus the contributions of the electromagnetic field allows us to obtain the  single-particle
dispersion relation between energy and momentum (the low velocities limit is being considered), as follows
\begin{eqnarray}
\label{SCE}
 E_{p}\simeq\frac{p^{2}}{2m}&+&\frac{\lambda}{2m} |\Phi|^{2}+\frac{\lambda}{4m}(k_B T)^{2}+m\phi+e \varphi\nonumber\\
&-&\frac{e}{m}\boldsymbol{A}\cdot \boldsymbol{p}
\end{eqnarray}

Unfortunately, because of the functional form of the scalar potential in Eq.(\ref{eq:potT}) and the plane wave \emph{ansatz}, the non-relativistic single-particle dispersion relation does not contain lower powers in the
temperature contributions caused by the thermal bath. However, it could be interesting to explore the existence of other scenarios, where lower power corrections caused by the thermal bath can be achieved. 

Nevertheless, as reference (Castellanos et al. \cite{CastMat}) mentions, large values of $\lambda$, which is function of the scattering length $a_s$, could be used to enlarge the contributions in the condensation temperature caused by the thermal bath, just tuning the interaction coupling by Feshbach resonances to large values of the scattering length, but where the diluteness condition,  $n|a_s|^{3}<<1$, remains valid.

The experimental realization of Bose-Einstein condensates has been achieved in experiments where the shape of the trapping potential is, in many cases, well approximated through a harmonic shape. For simplicity $\boldsymbol{A}$ is set equal to zero, and a dependence of the form $\varphi \sim r^{2}$ is taken for the electric potential. Clearly this can be generalized to other situations.

The spatial density associated with the system is given again by the following equation (Castellanos et al. \cite{CastMat})
\begin{equation}
\label{DE1}
n(\boldsymbol{r})=\Bigg(\frac{mk_{B}T}{2\pi\hbar^{2}}\Bigg)^{3/2}g_{3/2}(Z),
\end{equation}
but in this case, it is true that $Z$, as given by the following equation
\begin{equation}
Z=\exp [\beta(\mu-\frac{\lambda
\kappa^{-2}}{2m}n(\boldsymbol{r})-\frac{\lambda(k_{B}T)^{2}}{4m}-m\phi-e \varphi )]
\end{equation}
depends on the electric potential explicitly, being $g_{\nu}(z)$ the Bose-Einstein function (Pathria \cite{Pathria}).

In order to calculate the condensation temperature, Eq.(\ref{DE1}) is expanded up to first order in the coupling constant $\lambda$ using the properties of the Bose--Einstein functions
(Pathria \cite{Pathria}). With this at hand,
\begin{eqnarray}
\label{DE2} 
n(\boldsymbol{r}) \approx n_{0}(\boldsymbol{r})&-&\lambda g_{3/2}(z(\boldsymbol{r}))\Bigg[\frac{\Lambda^{-6} \kappa^{-2}}{2 mk_{B}T}g_{1/2}(z(\boldsymbol{r}))\nonumber\\
&+&\Lambda^{-3}\frac{k_{B}T}{4m}\frac{g_{1/2}(z(\boldsymbol{r}))}{g_{3/2}(z(\boldsymbol{r}))} \Bigg],
\end{eqnarray}
where $n_{0}(\boldsymbol{r})=\Lambda^{-3}g_{3/2}(z(\boldsymbol{r}))$ is the density for the case $\lambda=0$, being $\Lambda=(2\pi \hbar^{2}/mk_{B}T)^{1/2}$ the de Broglie thermal  wavelength and
\begin{equation}
 z(\boldsymbol{r})=\exp(\beta(\mu-\alpha mr^{2}-e \varphi )).
 \end{equation}

In order to calculate the number of particles composing the system, let us assume for simplicity that $\varphi \sim r^{2}$. Thus, by inserting the density of particles Eq.\,(\ref{DE2}) in the normalization condition $N=\int d^{3}\boldsymbol{r}\hspace{0.1cm}n(\boldsymbol{r})$, this allows us to obtain after integration the corresponding number of particles
\begin{eqnarray}
\label{NPR2} N &\simeq&
\Bigl(\frac{m}{2\Omega\hbar^{2}}\Bigr)^{3/2}(k_BT)^{3}g_{3}(\exp(\beta\mu))\nonumber\\
&-&\frac{\lambda\kappa^{-2}m^{2}(k_BT)^{7/2}}{16\pi^{3/2}\hbar^{6} \Omega^{3/2}}G_{3/2}(\exp(\beta
\mu))\nonumber\\
&-&\frac{\lambda}{4}\Bigl(\frac{m^{1/3}}{2\Omega\hbar^{2}}\Bigr)^{3/2}(k_BT)^{4
}g_{2}(\exp(\beta \mu)),
\end{eqnarray}
where $G_{3/2}(\exp(\beta \mu))=\sum_{i,j=1}^{\infty}\frac{\exp[(i+j)\beta \mu]}{i^{1/2}j^{3/2}(i+j)^{3/2}}$,
being  $\Omega=m(\alpha+const \times e)$. 

When $ \varphi $ is only position dependent, then from Eq. (\ref{DE2}) it can immediately be noticed how the correction in the number of particles can be associated with an effective external potential, and therefore $\Omega$  can be related to an effective frequency. 

If it is further assumed that above the condensation temperature the number of particles in the ground state is negligible, this allows us to obtain an expression for the condensation temperature $T_{0}$ in the non--interacting case, i.e., $\lambda=0$,
\begin{equation}
\label{CTI}
k_BT_{0}=\Bigl(\frac{2 \Omega \hbar^{2}
}{m}\Bigr)^{1/2}\Bigl(\frac{N}{\zeta(3)}\Bigr)^{1/3}.
\end{equation}

Additionally, at the condensation temperature, the chemical potential within the semiclassical
approximation can be expressed as $\mu_{c}=\frac{\lambda\kappa^{-2}}{2m}n(\boldsymbol{r}=0)$, such as Eq. (\ref{DE1}) suggests, thus
\begin{eqnarray}
\label{PQ1} \mu_{c}&\approx&\frac{\lambda\kappa^{-2}m^{1/2}(k_{B}T_{c})^{3/2}\zeta(3/2)}{2(2\pi)^{3/2}\hbar^{3}}\nonumber\\
&-&\lambda^{3/2}\frac{\sqrt{2}\pi\kappa^{-2}(k_{B}T_{c})^{2}}{(2\pi\hbar^{2})^{3/2}},
\end{eqnarray}
where $g_{3/2}(\exp(-\delta))\approx\zeta(3/2)-|\Gamma(-1/2)| \delta^{1/2}$ when $\delta\rightarrow 0$ has been used (Pathria \cite{Pathria}). Let us mention that Eq.\,(\ref{PQ1}) contains corrections  caused by the thermal bath which contributes with an extra term to the chemical potential at the transition temperature. The contributions of the thermal bath upon $\mu_{c}$ has as a consequence a shift in the calculation of the critical temperature when interactions are present.

By using these results, the shift in the condensation temperature caused by $\lambda$ and the thermal bath is finally obtained in function of the number of particles
\begin{eqnarray}
\label{CTR2}
\frac{T_{c}-T_{0}}{T_{0}} \equiv \frac{\Delta T_{c}}{T_{0}}=&-&\lambda\frac{m^{1/2}}{\kappa^{2}\hbar^{3}}\chi_{1}\Theta N^{1/6}\nonumber\\
&+&\lambda \chi_{2}\Theta^{2}\,N^{1/3}, \,\,\,\,\,\,\,\,\,
\end{eqnarray}
where 
\begin{eqnarray}
\chi_{1}&=&\frac{1}{3\zeta(3)}\left(\frac{\zeta(3/2)\zeta(2)}{2(2\pi)^{3/2}}-G_{3/2}(1)\right),\\ 
\chi_{2}&=&\frac{1}{3\zeta(3)}\left(\frac{1}{4\,mc^{2}}+\frac{(2\lambda)^{1/2}\zeta(2)\pi}{(2\pi)^{3/2}\kappa^{2}\hbar^{3}}\right),
\label{xi}
\end{eqnarray}
together with $\Theta=\left({2 \Omega\hbar^{2}}/{m}\right)^{1/4}$ and $T_{0}$ defined in Eq. (\ref{CTI}). The second term on the right hand side in the shift of Eq.(\ref{CTR2}) is the contribution due to the thermal bath and the field $\varphi$. 

Notice that if $\varphi=0$, the result given in reference (Castellanos et al. \cite{CastMat}) is recovered. By
setting $\alpha=1/2(\omega_{0})^{2}$ and $\lambda=8\pi \hbar^{2}\kappa^{2}a$ in Eq. (\ref{CTR2}), then the condensation temperature for a bosonic gas trapped in an isotropic harmonic oscillator is recovered, and this temperature is adjusted by the contributions of the thermal bath and the external field $\varphi $.

In other words, in order to have relevant adjustments over the usual result in typical laboratory conditions, the \emph{parameter} $\kappa$ must be very large and the external field $\varphi$ must be very weak, at least near to the center of the system. 

Thus, taking the experimental results for a $_{19}^{39}K$ condensate, the first  term on the right hand side of Eq. (\ref{CTR2}), which is produced by bosonic interactions, is of order $\sim 10^{-2}$ as expected (Smith et al. \cite{SM}). Additionally, notice that the order of magnitude for the second term in Eq.(\ref{CTR2}), is also an additive correction of the order of $10^{-2}$, for the same experimental conditions, where we have reintroduced $c=3\times10^{8}\,\mbox{meters}$\,$\mbox{seconds}^{-1}$. 
Since $\boldsymbol{A}$ is set equal to zero together with the assumption that $\varphi \sim r^{2}$ the above results must be taken carefully (for it is an approximated result).

\section{Conclusions}\label{sec:conclutions}

In this work the phase transition of a bosonic system with particle mass $m$ and self-interaction parameter $\lambda$ represented by the Klein-Gordon equation with a $U(1)$ symmetry at finite temperature with electromagnetic contributions was studied.

We have discussed Bose-Einstein condensation and other related systems at finite and zero temperature as well as its behavior when the application of electromagnetic fields is in order. A framework was setup that can be used to study the dynamics of the system at non zero temperature and with electromagnetic contributions. For instance, many of the classic results for the classical BEC at zero temperatures and zero applied electromagnetic fields where derived in an efficient and direct manner.

This model seems to have the capability of exhibiting symmetry breaking and Bose-Einstein condensation simultaneously but independently; from ordered to disordered phase depending on the value of the temperature above or below $T_c^{SB}$ and fulfilling Gross-Pitaevskii's equation for Bose-Einstein condensates once T=A=0 through the hydrodynamical representation. 

It was shown how the transition from the phase with the $U(1)$ symmetry to the phase with the broken symmetry can be related to the phase transition from the gas state to the condensation state of a Bose gas.

Again, at finite temperature significant changes are expected. First, as it was obtained in section \ref{sec:hydro}, the density of the system seems modified due to the thermal contribution. Second, the effects due to the self-interaction should be always taken into account since it characterizes the physical measurable properties of the gas, having a straightforward relation with the s-wave scattering length. 

And third, the quantum contribution of the quantum potential to the thermodynamics of the system was identified to be an intrinsic property of the system not depending on the initial and final states on which the equilibrium state can be reached.

Also, it was shown how only through the study of Eqs. (\ref{eq:hydro}) under the correct approximations it can be possible to relate different physical phenomena like Bose-Einstein condensation, superfluidity and superconductivity. All of them being described by the same system (set of equations) under different physical environments. 

The corresponding condensation temperature for the gas coupled to a electromagnetic field was calculated. In particular, leading quantum and electromagnetic corrections are derived through Klein-Gordon's equation for bosonic systems. A temperature and field dependent representation of the Klein-Gordon equation was obtained which can help to study the stability and behavior of the different statistical and quantum quantities involved as the temperature goes through the boundary of the phase transition, making a clear difference between condensed and non-condensed fractions within the system in the presence or absence of electromagnetic fields and thermal contributions.

This work summarizes our current understanding of some aspects in the connection between charged scalar field theory at finite temperatures and Bose-Einstein condensates at zero and finite temperatures with electromagnetic contributions. 

We hope the readers could be motivated in order that further research can be done and continued in the field. We believe that correct understanding of some theoretical problems is necessary for a more deep insight into experiments and their proper interpretation.

\acknowledgments 
This work was partially supported by CONACyT M\'exico
under grants CB-2009-01, no. 132400, CB-2011, no. 166212,  and I0101/131/07 C-
234/07 of the \emph{Instituto Avanzado de Cosmolog\'{\i}a} (IAC) collaboration
(http://www.iac.edu.mx/).  E. C.  and A. S. acknowledge
CONACyT for the postdoctoral grant received. E. C. acknowledges MCTP/UNACH for financial support.

\end{document}